\documentclass{Interspeech}

\interspeechcameraready

\usepackage[utf8]{inputenc} % allow utf-8 input
\usepackage[T1]{fontenc}    % use 8-bit T1 fonts
\usepackage{hyperref}       % hyperlinks
\usepackage{url}            % simple URL typesetting
\usepackage{booktabs}       % professional-quality tables
\usepackage{amsfonts}       % blackboard math symbols
\usepackage{nicefrac}       % compact symbols for 1/2, etc.
\usepackage{microtype}      % microtypography
\usepackage{xcolor}         % colors
\usepackage{cite} % [1, 2, 3, 4, 5] -> [1-5]
\hypersetup{colorlinks=true}
\usepackage[page]{appendix}
\usepackage{caption}
\usepackage{subcaption}
\usepackage{graphicx}
\usepackage{algorithm}
\usepackage[noend]{algpseudocode}
\usepackage{fancybox,framed}
\usepackage{multirow}
\usepackage{tabularx}
\usepackage{soul}
\usepackage{amsmath}
\usepackage{enumitem}
\usepackage{multirow}
\usepackage{diagbox}
\usepackage[export]{adjustbox}
\usepackage{array}
\usepackage{colortbl}
\newcommand\norm[1]{\left\lVert#1\right\rVert}
\usepackage{float}
\usepackage{tipa}
\usepackage[most]{tcolorbox}
\usepackage{makecell}
\usepackage{pifont}

\title{LLM-Synth4KWS: Scalable Automatic Generation and Synthesis of Confusable Data for Custom Keyword Spotting}

\author[]{Pai}{Zhu}
\author[]{Quan}{Wang}
\author[]{Dhruuv}{Agarwal}
\author[]{Kurt}{Partridge}

\affiliation[nocounter]{Google DeepMind}{New York, NY}{U.S.A}

\email{paizhu@google.com, quanw@google.com, dhruuv@google.com, kep@google.com}

\usepackage{comment}

\begin{document}

\maketitle

\begin{abstract}
    
    Custom keyword spotting (KWS) allows detecting user-defined spoken keywords from streaming audio. This is achieved by comparing the embeddings from voice enrollments and input audio. State-of-the-art custom KWS models are typically trained contrastively using  utterances whose keywords are randomly sampled from training dataset. These KWS models often struggle with confusing keywords, such as ``blue'' versus ``glue''. This paper introduces an effective way to augment the training with confusable utterances where keywords are generated and grouped from large language models (LLMs), and speech signals are synthesized with diverse speaking styles from text-to-speech (TTS) engines. To better measure user experience on confusable KWS, we define a new northstar metric using the average area under DET curve from confusable groups (c-AUC). Featuring high scalability and zero labor cost, the proposed method improves AUC by 3.7\% and c-AUC by 11.3\% on the Speech Commands testing set.
\end{abstract}

\section{Introduction}

The increasing popularity of voice-controlled devices including smart phones, wearables and home devices necessitates high-performance keyword spotting (KWS) models with low power consumption and memory footprint. Conventional KWS approaches~\cite{alvarez2019end, labrador2023personalizingkeywordspottingspeaker, zhu2023locale} focus on detecting pre-defined keywords (e.g. ``Alexa'', ``Hey Google''). However, the rising demand for personalization in smart devices has spurred the need for custom KWS solutions, enabling users to select triggering keywords from an open vocabulary based on their personal preference for device and software activation.

Recently, representation learning approaches achieved state-of-the-art (SOTA) results in custom KWS tasks~\cite{ding2022letr, lin2020training, rybakov2020streaming}, where speech utterances are represented by fixed-length vectors, and the similarity between two vectors is compared to a pre-defined threshold to make detection decisions. 
Contrastive learning was then proposed to improve KWS accuracy, leveraging loss functions such as triplet loss~\cite{sacchi2019open,chidhambararajan2022efficientword} and generalized end-to-end (GE2E) loss~\cite{zhu2024ge2ekwsgeneralizedendtoendtraining}.
However, both triplet and GE2E approaches sample contrastive utterances from keywords that are randomly chosen from the training vocabulary. The chance of grouping confusable keywords in the same training batch is very low. While those models perform decently in differentiating distinct pronunciations (e.g. ``blue'' versus ``goat''), they struggle with similar-sounding keywords such as ``blue'' versus ``glue''. 

To optimize the model's performance on confusable keyword spotting,
past work~\cite{gao2020dataefficientmodelingwakeword, kewei2024phonemelevelcontrastivelearninguserdefined} use Levenshtein distance to perform confusable example search inside the training dataset to build contrastive utterance pairs for each training batch. While improving confusable utterance detections, the search time complexity $O(N)$ at each training batch significantly slows down the training process. A similar approach is also used in custom KWS for Chinese keywords~\cite{5597890}. Although they maintain a confusable example lookup table to expedite the training, the runtime memory overhead makes the approach resource-intensive and unscalable.

\begin{figure*}[!h]
	\centering
	\includegraphics[width=1\textwidth]{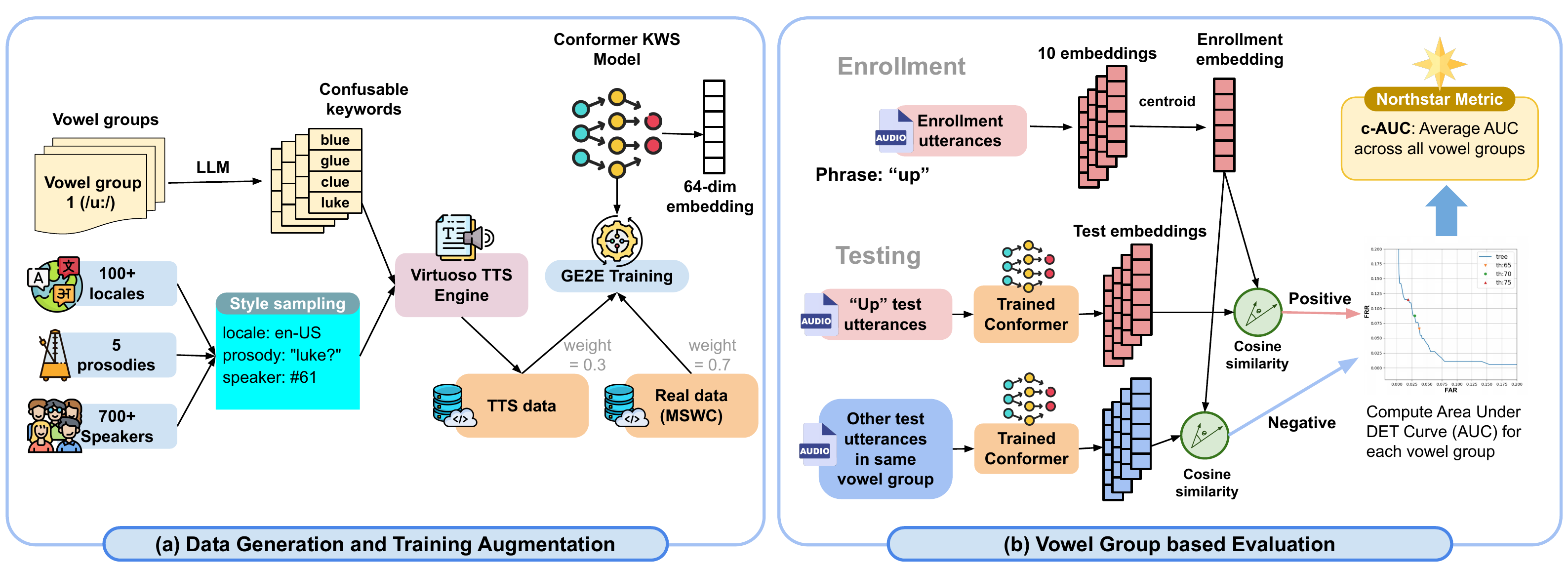}
	\caption{An overview of the LLM-Synth4Kws framework: (a) LLM generating confusable keywords, style sampling for TTS, and training data augmentation. (b) Vowel group based evaluation and a new northstar metric.}
	\label{fig:ab_graph}
	\vspace{-5mm}
\end{figure*}

The recent breakthroughs of Large Language Models (LLMs) revolutionize AI applications especially in text generation tasks. LLMs have also been applied in the speech domain to improve speech recognition accuracy and enable context-aware ability~\cite{bai2024seedasrunderstandingdiversespeech}. Meanwhile, with recent advances of Text-to-Speech (TTS) technologies including MAESTRO~\cite{chen2022maestro} and Virtuoso~\cite{saeki2023virtuoso}, speech utterances can be synthesized with high fidelity from a wide distribution of speaking characteristics.
TTS data augmentation has been widely used in speech recognition~\cite{wang2020improving}, speaker verification~\cite{huang2021synth2aug}, and improving KWS accuracy in low-resource environments~\cite{zhu2024synth4kws,ttsforkws2}. Particularly, TTS has been explored in improving confusable keyword detection in~\cite{jia2020trainingwakeworddetection, wang2022generatingadversarialsamplestraining,graphemeaug}. However those approaches focus on multi-speaker speech synthesis where augmenting utterances are sampled from 7000 speaker embeddings based on only 12 manually-created confusing words in Chinese language. The insufficiency of word diversity limits the model's generalization power for confusable KWS tasks.

The main contributions of this paper include:
\setlist{nolistsep}
\begin{itemize}[leftmargin=*]
\item \textbf{A new northstar metric}: Our user study suggests that custom KWS models particularly struggle to differentiate words that share vowels, which is not reflected in SOTA evaluation metrics. To directly measure confusable KWS performance, we propose a vowel group based metric denoted as \textbf{c-AUC}.
\item  \textbf{Data augmentation}: We design a scalable and automatic framework to generate group based confusable keywords by LLM prompting, and synthesize utterances with diverse accents and prosodies from TTS.
\item \textbf{Training schema}: We create an efficient training schema by adapting GE2E batch to sample vowel grouped examples and optimize contrastive loss between confusable utterances.
\end{itemize}

In addition to significant performance improvements, our approaches have the following advantages:
\setlist{nolistsep}
\begin{itemize}[leftmargin=*]
\item {\bf Training efficiency}. Our augmentation method does not need to search for similar words during training nor maintain an expensive word lookup table. 
\item {\bf Scalability}. Today's LLM and TTS support many languages with great quality. This allows us to improve custom KWS in many other languages, especially resource-constrained ones.
\item {\bf Affordability and availability}. While crowdsourcing projects take months, LLM and TTS can generate large volumes of data tailored to on-demand task requirements in hours and with a very low cost.
\end{itemize}

\section{Methods}
\label{sec:methods}

\subsection{Baseline Model: Architecture and GE2E Loss}

Our baseline model is a 420KB quantized conformer~\cite{gulati2020conformer} model  optimized for always-on devices.
During training, we use Generalized End-to-End loss (GE2E)~\cite{wan2018generalized} for its performance and computation advantages over the traditional triplet loss~\cite{sacchi2019open}, and tailor it for the KWS task~\cite{zhu2024ge2ekwsgeneralizedendtoendtraining}. Instead of sampling one positive example pair and one negative example pair per batch in the triplet method, we construct a batch with $X$ phrases and $Y$ sampled utterances per phrase. For each phrase, we used half the sampled utterances $Y/2$ to build the enrollment embedding centroid, and compare it with the embeddings of other $Y/2$ test utterances using cosine similarity. The usage of enrollment centroid and diverse test utterances in each training batch greatly reduces loss variance, producing more stable convergence and better accuracy.

\subsection{Generating Confusable Keyword Groups using LLM}
\label{subsec:llm}

The current training batch randomly chooses $X$ words from the whole vocabulary of training dataset $\mathbf{V}$. 
Assuming $\norm{\mathbf{V}} \gg X$,  the chance of grouping confusable keyword utterances in a same training batch is very low. This causes high False Accepts (FAs) in production systems, especially when the enrollment utterance contains words with few syllables. 

\begin{figure}[!ht]
	\centering
	\includegraphics[width=0.32\textwidth]{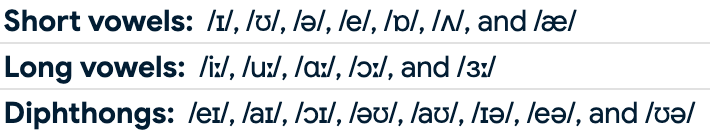}
	\vspace{-2mm}
	\caption{Vowels in the English language.}
	\label{fig:vowel_groups}
	\vspace{-5mm}
\end{figure}

We tackle this problem by augmenting training with example batches containing only confusable keywords. Our live demo testing and user feedback suggest that KWS models particularly struggle to differentiate words that share vowels in the English language. We therefore design groups around confusable words with shared vowels. There are 20 vowels in English~\cite{vowels} as shown in Figure~\ref{fig:vowel_groups}. Illustrated in Figure~\ref{fig:ab_graph}(a), for each vowel, we prompt the LLM to create a list of words containing the corresponding vowel. For example, the prompt to generate vowel \textipa{[u:]} grouped keywords is:

\scalebox{0.95}{
\begin{tcolorbox}
[enhanced,attach boxed title to top center={yshift=-3mm,yshifttext=-1mm}, title=\textbf{Prompt for single vowel}, colback=blue!3!white, colframe=blue!50!white, coltitle=black, colbacktitle=blue!10!white]
Can you generate a group of words? The group should have 100 simple but distinguishable words. All words in the group should contain the vowel \textipa{[u:]}
\end{tcolorbox}}

\begin{figure}[!ht]
	\centering
	\includegraphics[width=0.32\textwidth]{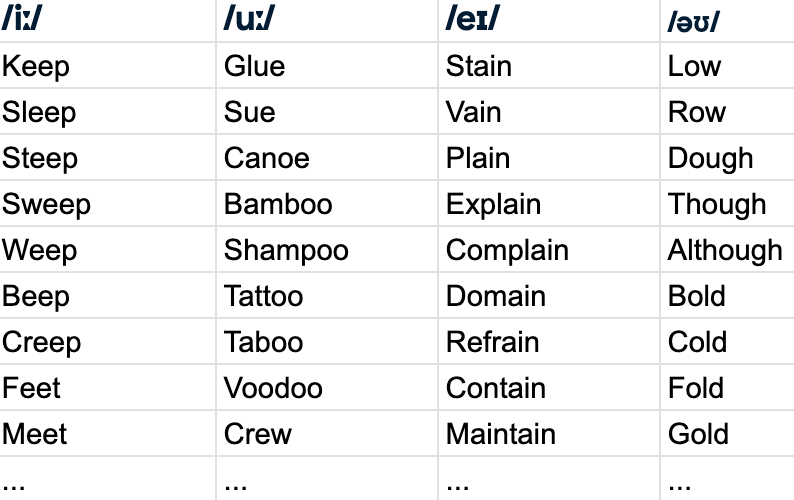}
	\vspace{-2mm}
	\caption{Vowel grouped words generated by LLM (truncated).}
	\label{fig:group_words}
	\vspace{-5mm}
\end{figure}

We repeat the above prompt for all 20 vowels. Figure~\ref{fig:group_words} shows the partial results of LLM generated words in different groups. In this process, we prompt Gemini 1.5 Pro~\cite{geminiteam2024gemini15unlockingmultimodal} via Google AI Studio to avoid generating complicated or long words, so the model can focus on the differences between similar-sounding keywords.

\subsection{TTS Style Sampling}
\label{subsec:tts}

The Virtuoso TTS engine~\cite{saeki2023virtuoso} is well known for its naturalness, intelligibility, prosody control and language coverage (139 locales). While our experiments focus on English, this approach easily generalizes to other languages, even those with limited real world data.

As shown in Figure~\ref{fig:ab_graph}(a), for each keyword generated by LLM from different groups, we use Virtuoso to generate 100 utterances, sampling from 726 speakers with diverse demographics and accents and five prosodies. Meanwhile, Virtuoso allows to control utterance prosody by using punctuation symbols, and we use the same setup as ~\cite{park2024utilizingttssynthesizeddata} to augment the prosody of the output.  Therefore, we create the augmented dataset based on TTS synthesized utterances from LLM generated keywords.

\subsection{LLM/TTS Dataset and Training Augmentation}
As shown in Figure~\ref{fig:ab_graph}(a), we augment model training by using the above-mentioned LLM and TTS generated confusable dataset (referred to as LLM/TTS dataset henceforth). Specifically, in each training batch, all examples are drawn from either the LLM/TTS dataset or the Multilingual Spoken Words Corpus (MSWC)~\cite{mazumder2021multilingual} with their respective probabilities $P_{mswc}$ and $P_{tts}$.

\section{Data and Experiment Setup}
\label{sec:data}

\subsection{Training Resources}
\label{subsec:training_resource}
The training data for the baseline model are from the MSWC dataset\cite{mazumder2021multilingual}, consisting of 5.3M utterances from 38k phrases and 27k speakers. For each  utterance, we extract 40-dimensional log-Mel filterbank energies from 25ms frames.

Our models are developed using the Tensorflow/Lingvo~\cite{Shen2019LingvoAM} framework. Dynamic range quantization is used to quantize weights and activations to 8-bit precision integers.

\subsection{Experiment Setup}
\label{subsec:experiment_setup}
For the baseline model, we construct the training batch by sampling $X=8$ keywords from the entire MSWC dataset, and then sample $Y=10$ utterances per keyword. Each training batch contains 80 utterances.

We use a similar process to sample the LLM/TTS-augmented dataset, but the utterances from the same vowel group are used to construct the training batch. For each vowel group, we randomly choose $X=8$ keywords and then randomly choose $Y=10$ utterances per keyword.

For the proposed model, each training batch samples all utterances from either MSWC dataset or LLM/TTS dataset, with probability $P_{mswc}=0.5$ and probability $P_{tts}=0.5$ respectively.

\subsection{Evaluation Datasets and Process}

We use the test split of Speech Commands dataset~\cite{warden2018speech} and a subset of LibriPhrase~\cite{shin2022learningaudiotextagreementopenvocabulary} for evaluation purposes.

Speech Commands testing set comprises over 11,000 utterances spanning 35 distinct phrases. As shown in Figure~\ref{fig:ab_graph}(b), for each testing phrase (e.g. ``up"), 10 randomly selected utterances are used as the enrollment set, with the remaining designated for testing.

LibriPhrase consists of
39k positive pairs from the same keywords;
39k easy negative pairs from different keywords (such as ``apartment'' v.s. ``water'');
and 39k hard negative pairs from different keywords (such as ``apartment'' v.s. ``abatement'').
We take a subset of these utterances (denoted as \textit{LibriPhrase-1s}) whose durations are between $0.9 \sim 1.1$ second as our evaluation set. This subset of LibriPhrase is more consistent with our training configuration (1 second).

\subsection{Evaluation Metrics}
\label{subsec:eval_metrics}

\subsubsection{Speech Commands metrics: EER and AUC}
\label{subsubsec:full_metrics}
After computing the cosine similarity score between the enrollment embedding and test utterance embedding, we compare it with a threshold swept from 0 to 1 with an increment of 0.01. This allows us to plot the Detection Error Tradeoff (DET) curve for the False Accept Rates (FARs) and False Reject Rates (FRRs). Two threshold-independent metrics are computed to evaluate model performance:
\begin{itemize}[leftmargin=*]
\itemsep0em
\item \textbf{Equal Error Rate (EER)}: The FAR value at the operating point where FAR and FRR are equal. Smaller is better.
\item \textbf{Area Under the Curve (AUC)}: The area under the DET curve. Smaller is better.
\end{itemize}
To measure model performance across different keywords, the EER and AUC metrics are aggregated from 35 keywords in the Speech Commands dataset.

\subsubsection{Speech Commands metrics: c-AUC}
In order to further analyze how LLM/TTS-augmentation improves KWS specifically for confusable keywords, we group distinct keywords in Speech Commands dataset based on the vowels as shown in Table~\ref{tab:vow_def}, with the following rules:
\begin{itemize}[leftmargin=*]
\itemsep0em
\item Very similar vowels such as \textipa{i:} and \textipa{I} will be grouped together.
\item A group must have at least 3 words otherwise will be discarded.
\end{itemize}

Unlike the metrics in Section~\ref{subsubsec:full_metrics} where the enrollment is compared with test utterances from all 35 keywords in Speech Commands dataset, in the group based evaluation, the enrollment is only compared with the test utterances from the same vowel group. For example, the enrollment keyword `right' will be only compared with test utterances from /\textipa{aI}/ group, i.e. `nine' and `five'. For each group we compute vowel group AUC using the same method as Section~\ref{subsubsec:full_metrics}.

We average vowel group AUCs from all defined groups in Table~\ref{tab:vow_def}, and refer it as \textbf{confusable group based AUC (c-AUC)}. This metric measures KWS performance on detecting confusable utterances within their vowel group.

\begin{table}[!ht]
\centering
\caption{Vowel groups and evaluation keywords from Speech Commands.}
\resizebox{6cm}{!}{%
\begin{tabular}{l|l}
\toprule
 \textbf{Vowel Group} & \textbf{Words} \\ \midrule
\textipa{/a:/} & `marvin', `on', `stop'      \\
\textipa{/o:/} & `dog', `four', `off' \\
\textipa{/e/}: & `yes', `seven', `left', `bed' \\
\textipa{/\ae/} & `cat', `backward', `happy' \\
\textipa{/i:/} and \textipa{/I/} & `sheila', `tree', `three', `visual', `six'     \\
\textipa{/E:/} and \textipa{/@/} & `forward', `bird', `learn'   \\
\textipa{/aI/} & `right', `nine', `five' \\
\textipa{/aU/} & `wow', `down', `house' \\
\textipa{/oU:/} and \textipa{/u:/} & `go', `no', `follow', `zero', `two' \\
\bottomrule
\end{tabular}}
\label{tab:vow_def}
\vspace{-5mm}
\end{table}

\subsubsection{LirbPhrase-1s metrics}
For LibriPhrase-1s, we plot two DET curves: combining positive pairs and easy negative pairs; and combining positive pairs and hard negative pairs. The resulting AUC metrics will be referred to as Easy-AUC and Hard-AUC respectively. The Hard-AUC metrics for LibriPhrase-1s plays a similar role to the c-AUC metric we defined for the Speech Commands dataset.

\section{Experimental Results}
\label{sec:results}

The main experimental results are presented in Table~\ref{tab:eval_results}.
The baseline model is trained on MSWC dataset as described in Section~\ref{subsec:experiment_setup}. The proposed approach, denoted as (\textbf{AugModel}) is trained on both MSWC and LLM/TTS synthetic data, where LLM is used to generate vowel grouped confusable keywords as discussed in Section~\ref{subsec:llm},  and TTS is used to synthesize utterances as discussed in Section~\ref{subsec:tts}.

\subsection{Results on Speech Commands}
\label{subsec:results1}

\begin{table*}[]
\centering
\caption{Evaluation results for the baseline and LLM/TTS-augmented model on Speech Commands and LibriPhrase-1s.}
\vspace{-2mm}
\begin{tabular}{c| c c |c c c|c c}
\toprule
& \multicolumn{2}{c|}{Training Data} & \multicolumn{3}{c|}{Speech Commands Testing Set} & \multicolumn{2}{c}{LibriPhrase-1s} \\
 & MSWC  & LLM/TTS  & EER (\%) & AUC (\%) &  c-AUC (\%) &  Easy-AUC (\%) & Hard-AUC (\%) \\ \midrule
Baseline Model     & \ding{52} &     \ding{56}          & 2.49                 & 0.490  & 0.915 & 0.012 & 14.4   \\
AugModel       & \ding{52} &    \ding{52}        & 2.38                 & 0.472    & 0.812 & 0.0 & 12.6 \\ \midrule
\textbf{Improvement (relative)}     & - & -     & \textbf{4.4}         & \textbf{3.7}   & \textbf{11.3}   & \textbf{100.0} & \textbf{12.5}  \\
\bottomrule
\end{tabular}
\label{tab:eval_results}
\vspace{-5mm}
\end{table*}

From Table~\ref{tab:eval_results}, our augmented model achieves relative reductions of 4.4\% in EER and 3.7\% in AUC on Speech Commands. This is mainly due to the confusable data augmentation enhanced the model's ability to differentiate similar-sounding keywords inside Speech Commands dataset. However the improvement is relatively marginal, because in most cases, the enrollment is compared with test utterances from non-confusable keywords, which is not directly optimized by the augmentation.

\begin{figure}[!ht]
	\centering
	\includegraphics[width=0.4\textwidth]{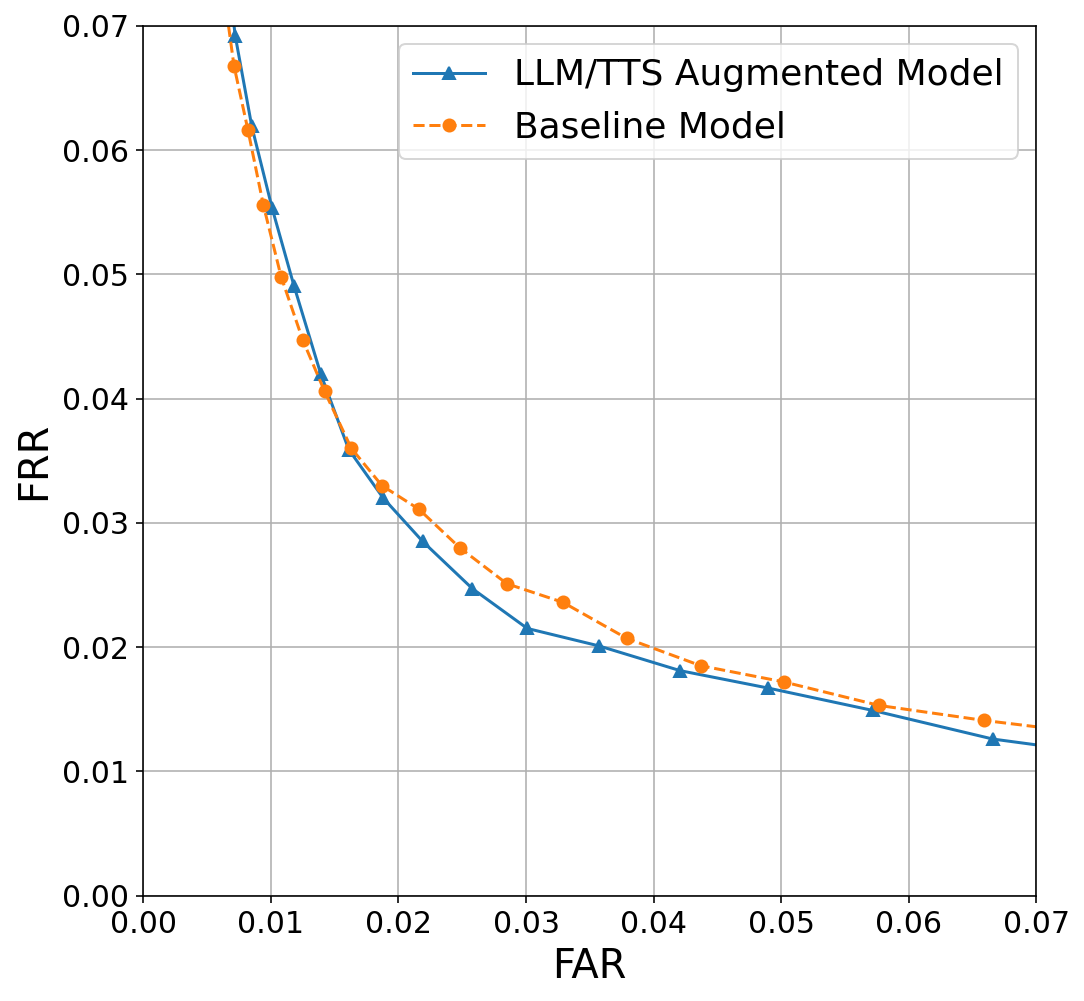}
	\vspace{-3mm}
	\caption{DET curves for baseline and augmented model on Speech Commands.}
	\label{fig:det_results}
	\vspace{-5mm}
\end{figure}

The DET curves in Figure~\ref{fig:det_results} show the LLM/TTS-augmented model performs better than baseline model at most operating points. However, at the very low FAR side (i.e. FAR < 1\%), when the matching threshold is much higher, the augmented model performs slightly worse. One possible reason is that TTS data introduced bias towards the synthesized voice, and the shifted prior distribution makes the model less confident about the real utterances. At a high matching threshold, this results in high FRRs when evaluating real utterances from Speech Commands dataset. As future work, we will experiment with adversarial training method~\cite{park2024adversarialtrainingkeywordspotting} to minimize TTS overfitting.

\subsection{Results on Confusable Keywords}

For each vowel group defined in Table~\ref{tab:vow_def}, Table~\ref{tab:result_group_based} shows the averaged AUC over different words in that group. Comparing to the AUC results in Table~\ref{tab:eval_results} which is based on the entire Speech Commands dataset, we can see the absolute values of group based AUC are mostly higher, as the evaluation focuses on more challenging KWS tasks --- detecting confusable utterances inside vowel groups. However we can also see the LLM/TTS-augmented model produces much more significant improvements over baseline model for most confusable groups. Averaging the results from 9 different confusable groups, we can see the confusable group average: \textbf{c-AUC is improved by 11.3\%}.

The improvements vary across vowel groups due to the limitation of keyword diversity (3-5 distinct keywords per group) after grouping Speech Commands dataset. While most vowel groups see significant improvements from the LLM/TTS-augmentation,
the group /\textipa{i:}/ and /\textipa{I}/ has a slight degradation. Investigation indicates it is mostly caused by utterances from `three' and `tree' keywords. 
Since they both exist in the augmented dataset, one potential reason is TTS produces near-identical sounds for keyword `three' and `tree', and our training still forces the model to differentiate them. As future work, we will explore a TTS system that synthesizes more clear and distinct sounds among very similar words.

\begin{table}[!ht]
\centering
\caption{Vowel group based metrics for confusable keywords on Speech Commands.}
\vspace{-2mm}
\begin{tabular}{c|c|c|c}
\toprule
\textbf{Group}                 & \bfseries\makecell{  Baseline  \\ AUC (\%)} & \bfseries\makecell{  AugModel  \\ AUC (\%)} & \bfseries\makecell{  Relative  \\ Improv. (\%)} \\ \cline{1-4}
\textipa{a:}                  & 0.227              & 0.163                 & 28.15\%        \\ \cline{1-4}
\textipa{o:}                  & 1.044              & 0.636                 & 39.10\%         \\ \cline{1-4}
\textipa{e}                   & 0.394              & 0.382                 & 2.89\%         \\ \cline{1-4}
\textipa{\ae}  & 1.147              & 1.112                 & 3.12\%                   \\ \cline{1-4}
\textipa{\i:} and \textipa{I} & 2.221              & 2.435                 & -9.67\%        \\ \cline{1-4}
\textipa{E:} and \textipa{@} & 1.136              & 1.007                 & 11.37\%        \\ \cline{1-4}
\textipa{aI}                  & 0.873              & 0.583                 & 33.23\%        \\ \cline{1-4}
\textipa{aU}                  & 0.727              & 0.528                 & 27.35\%        \\ \cline{1-4}
\textipa{oU} and \textipa{u:} & 0.466              & 0.466                 & 0.19\%         \\ \cline{1-4}
\textbf{Average}                        & \textbf{0.915}              & \textbf{0.812}                 &  \textbf{11.3\%}    \\
\bottomrule
\end{tabular}
\label{tab:result_group_based}
\vspace{-5mm}
\end{table}

\subsection{Results on LibriPhrase-1s}

In Table \ref{tab:eval_results} we also present results on LibriPhrase-1s. Our augmented model improves AUC from 0.012\% to 0.0\% (no errors) for easy negative pairs, and improves AUC from 14.4\% to 12.6\% (12.5\% relative improvement) for hard negative pairs. We can see our LLM/TTS-augmented model significantly enhanced the model's ability to detect confusable keywords, which is consistent with the results on Speech Commands.

However, we would also like to point out that when evaluating on the entire LibriPhrase dataset with a broader range of utterance durations (with a mean of 0.63s and a standard deviation of 0.11s), the baseline model's performance largely degraded, and we did not observe improved metrics from the augmented model. This indicates the importance of consistency between training and inference sequence length. As a future work, we will retrain the model with LibriPhrase training set to improve model quality on various utterance lengths.

\section{Conclusion}
\label{sec:conclusion}

In this paper, we presented a novel approach to enhance the robustness of custom keyword spotting (KWS) systems, particularly for confusable keywords. We designed a scalable and automatic method to augment the training process with confusable keywords generated from LLMs and diverse utterances sampled from TTS. Our proposed model improves AUC by 3.7\% for  KWS task on the Speech Commands dataset. Moreover, we propose c-AUC as a novel northstar metric to bridge the real-world user experience and directly measure KWS performance over confusable utterances, and our proposed model improves corresponding c-AUC by 11.3\%. On LibriPhrase-1s, our proposed model also improves Hard-AUC by 12.5\%.
While our experiments focus on English, the proposed approach can be easily extended to 100+ other languages through the multilingual capabilities of LLMs and TTS data synthesis.

\clearpage
\bibliographystyle{IEEEtran}
\bibliography{mybib}

\begin{thebibliography}{10}
\providecommand{\url}[1]{#1}
\csname url@samestyle\endcsname
\providecommand{\newblock}{\relax}
\providecommand{\bibinfo}[2]{#2}
\providecommand{\BIBentrySTDinterwordspacing}{\spaceskip=0pt\relax}
\providecommand{\BIBentryALTinterwordstretchfactor}{4}
\providecommand{\BIBentryALTinterwordspacing}{\spaceskip=\fontdimen2\font plus
\BIBentryALTinterwordstretchfactor\fontdimen3\font minus
  \fontdimen4\font\relax}
\providecommand{\BIBforeignlanguage}[2]{{%
\expandafter\ifx\csname l@#1\endcsname\relax
\typeout{** WARNING: IEEEtran.bst: No hyphenation pattern has been}%
\typeout{** loaded for the language `#1'. Using the pattern for}%
\typeout{** the default language instead.}%
\else
\language=\csname l@#1\endcsname
\fi
#2}}
\providecommand{\BIBdecl}{\relax}
\BIBdecl

\bibitem{alvarez2019end}
R.~Alvarez and H.-J. Park, ``End-to-end streaming keyword spotting,'' in
  \emph{ICASSP 2019-2019 IEEE International Conference on Acoustics, Speech and
  Signal Processing (ICASSP)}.\hskip 1em plus 0.5em minus 0.4em\relax IEEE,
  2019, pp. 6336--6340.

\bibitem{labrador2023personalizingkeywordspottingspeaker}
B.~Labrador, P.~Zhu, G.~Zhao, A.~S. Scarpati, Q.~Wang, A.~Lozano-Diez, and
  I.~Lopez-Moreno, ``Personalizing keyword spotting with speaker information,''
  in \emph{ICASSP 2025-2025 IEEE International Conference on Acoustics, Speech
  and Signal Processing (ICASSP)}.\hskip 1em plus 0.5em minus 0.4em\relax IEEE,
  2025, pp. 1--5.

\bibitem{zhu2023locale}
P.~Zhu, H.~J. Park, A.~Park, A.~S. Scarpati, and I.~L. Moreno, ``Locale
  encoding for scalable multilingual keyword spotting models,'' in \emph{ICASSP
  2023-2023 IEEE International Conference on Acoustics, Speech and Signal
  Processing (ICASSP)}.\hskip 1em plus 0.5em minus 0.4em\relax IEEE, 2023, pp.
  1--5.

\bibitem{ding2022letr}
K.~Ding, M.~Zong, J.~Li, and B.~Li, ``Letr: A lightweight and efficient
  transformer for keyword spotting,'' in \emph{ICASSP 2022-2022 IEEE
  International Conference on Acoustics, Speech and Signal Processing
  (ICASSP)}.\hskip 1em plus 0.5em minus 0.4em\relax IEEE, 2022, pp. 7987--7991.

\bibitem{lin2020training}
J.~Lin, K.~Kilgour, D.~Roblek, and M.~Sharifi, ``Training keyword spotters with
  limited and synthesized speech data,'' in \emph{ICASSP 2020-2020 IEEE
  International Conference on Acoustics, Speech and Signal Processing
  (ICASSP)}.\hskip 1em plus 0.5em minus 0.4em\relax IEEE, 2020, pp. 7474--7478.

\bibitem{rybakov2020streaming}
O.~Rybakov, N.~Kononenko, N.~Subrahmanya, M.~Visontai, and S.~Laurenzo,
  ``Streaming keyword spotting on mobile devices,'' \emph{arXiv preprint
  arXiv:2005.06720}, 2020.

\bibitem{sacchi2019open}
N.~Sacchi, A.~Nanchen, M.~Jaggi, and M.~Cernak, ``Open-vocabulary keyword
  spotting with audio and text embeddings,'' in \emph{Prod. Interspeech}, 2019.

\bibitem{chidhambararajan2022efficientword}
R.~Chidhambararajan, A.~Rangapur, S.~Sibi~Chakkaravarthy, A.~K. Cherukuri,
  M.~V. Cruz, and S.~S. Ilango, ``{EfficientWord-Net}: An open source hotword
  detection engine based on few-shot learning,'' \emph{Journal of Information
  \& Knowledge Management}, vol.~21, no.~04, p. 2250059, 2022.

\bibitem{zhu2024ge2ekwsgeneralizedendtoendtraining}
P.~Zhu, J.~W. Bartel, D.~Agarwal, K.~Partridge, H.~J. Park, and Q.~Wang,
  ``{GE2E-KWS}: Generalized end-to-end training and evaluation for zero-shot
  keyword spotting,'' in \emph{2024 IEEE Spoken Language Technology Workshop
  (SLT)}.\hskip 1em plus 0.5em minus 0.4em\relax IEEE, 2024, pp. 999--1006.

\bibitem{gao2020dataefficientmodelingwakeword}
Y.~Gao, Y.~Mishchenko, A.~Shah, S.~Matsoukas, and S.~Vitaladevuni, ``Towards
  data-efficient modeling for wake word spotting,'' in \emph{ICASSP 2020-2020
  IEEE International Conference on Acoustics, Speech and Signal Processing
  (ICASSP)}.\hskip 1em plus 0.5em minus 0.4em\relax IEEE, 2020, pp. 7479--7483.

\bibitem{kewei2024phonemelevelcontrastivelearninguserdefined}
L.~Kewei, Z.~Hengshun, S.~Kai, D.~Yusheng, and D.~Jun, ``Phoneme-level
  contrastive learning for user-defined keyword spotting with flexible
  enrollment,'' \emph{arXiv preprint arXiv:2412.20805}, 2024.

\bibitem{5597890}
S.~Zhang, Z.~Shuang, Q.~Shi, and Y.~Qin, ``Improved mandarin keyword spotting
  using confusion garbage model,'' in \emph{2010 20th International Conference
  on Pattern Recognition}, 2010, pp. 3700--3703.

\bibitem{bai2024seedasrunderstandingdiversespeech}
Y.~Bai, J.~Chen, J.~Chen, W.~Chen, Z.~Chen, C.~Ding, L.~Dong, Q.~Dong, Y.~Du,
  K.~Gao \emph{et~al.}, ``{Seed-ASR}: Understanding diverse speech and contexts
  with llm-based speech recognition,'' \emph{arXiv preprint arXiv:2407.04675},
  2024.

\bibitem{chen2022maestro}
Z.~Chen, Y.~Zhang, A.~Rosenberg, B.~Ramabhadran, P.~Moreno, A.~Bapna, and
  H.~Zen, ``Maestro: Matched speech text representations through modality
  matching,'' \emph{arXiv preprint arXiv:2204.03409}, 2022.

\bibitem{saeki2023virtuoso}
T.~Saeki, H.~Zen, Z.~Chen, N.~Morioka, G.~Wang, Y.~Zhang, A.~Bapna,
  A.~Rosenberg, and B.~Ramabhadran, ``Virtuoso: Massive multilingual
  speech-text joint semi-supervised learning for text-to-speech,'' in
  \emph{ICASSP 2023-2023 IEEE International Conference on Acoustics, Speech and
  Signal Processing (ICASSP)}.\hskip 1em plus 0.5em minus 0.4em\relax IEEE,
  2023, pp. 1--5.

\bibitem{wang2020improving}
G.~Wang, A.~Rosenberg, Z.~Chen, Y.~Zhang, B.~Ramabhadran, Y.~Wu, and P.~Moreno,
  ``Improving speech recognition using consistent predictions on synthesized
  speech,'' in \emph{ICASSP 2020-2020 IEEE International Conference on
  Acoustics, Speech and Signal Processing (ICASSP)}.\hskip 1em plus 0.5em minus
  0.4em\relax IEEE, 2020, pp. 7029--7033.

\bibitem{huang2021synth2aug}
Y.~Huang, Y.~Chen, J.~Pelecanos, and Q.~Wang, ``{Synth2Aug}: Cross-domain
  speaker recognition with {TTS} synthesized speech,'' in \emph{2021 IEEE
  Spoken Language Technology Workshop (SLT)}.\hskip 1em plus 0.5em minus
  0.4em\relax IEEE, 2021, pp. 316--322.

\bibitem{zhu2024synth4kws}
P.~Zhu, D.~Agarwal, J.~W. Bartel, K.~Partridge, H.~J. Park, and Q.~Wang,
  ``{Synth4Kws}: Synthesized speech for user defined keyword spotting in low
  resource environments,'' \emph{arXiv preprint arXiv:2407.16840}, 2024.

\bibitem{ttsforkws2}
A.~Werchniak, R.~B. Chicote, Y.~Mishchenko, J.~Droppo, J.~Condal, P.~Liu, and
  A.~Shah, ``Exploring the application of synthetic audio in training keyword
  spotters,'' in \emph{ICASSP 2021 - 2021 IEEE International Conference on
  Acoustics, Speech and Signal Processing (ICASSP)}, 2021, pp. 7993--7996.

\bibitem{jia2020trainingwakeworddetection}
Y.~Jia, Z.~Cai, M.~Ma, Z.~Zhao, X.~Wang, J.~Wang, and M.~Li, ``Training wake
  word detection with synthesized speech data on confusion words,'' \emph{arXiv
  preprint arXiv:2011.01460}, 2020.

\bibitem{wang2022generatingadversarialsamplestraining}
H.~Wang, Y.~Jia, Z.~Zhao, X.~Wang, J.~Wang, and M.~Li, ``Generating adversarial
  samples for training wake-up word detection systems against confusing
  words,'' \emph{arXiv preprint arXiv:2201.00167}, 2022.

\bibitem{graphemeaug}
H.~Zhang, K.~Partridge, P.~Zhu, N.~Chen, H.~J. Park, D.~Agarwal, and Q.~Wang,
  ``{GraphemeAug}: A systematic approach to synthesized hard negative keyword
  spotting examples,'' \emph{arXiv preprint arXiv:2505.14814}, 2025.

\bibitem{gulati2020conformer}
A.~Gulati, J.~Qin, C.-C. Chiu, N.~Parmar, Y.~Zhang, J.~Yu, W.~Han, S.~Wang,
  Z.~Zhang, Y.~Wu \emph{et~al.}, ``Conformer: Convolution-augmented transformer
  for speech recognition,'' \emph{arXiv preprint arXiv:2005.08100}, 2020.

\bibitem{wan2018generalized}
L.~Wan, Q.~Wang, A.~Papir, and I.~L. Moreno, ``Generalized end-to-end loss for
  speaker verification,'' in \emph{2018 IEEE International Conference on
  Acoustics, Speech and Signal Processing (ICASSP)}.\hskip 1em plus 0.5em minus
  0.4em\relax IEEE, 2018, pp. 4879--4883.

\bibitem{vowels}
\BIBentryALTinterwordspacing
``Vowels in {English},'' accessed: 2025-01-10. [Online]. Available:
  \url{https://corpus.eduhk.hk/english\_pronunciation/index.php/2-1-english-vowels/}
\BIBentrySTDinterwordspacing

\bibitem{geminiteam2024gemini15unlockingmultimodal}
G.~Team, P.~Georgiev, V.~I. Lei, R.~Burnell, L.~Bai, A.~Gulati, G.~Tanzer,
  D.~Vincent, Z.~Pan, S.~Wang \emph{et~al.}, ``Gemini 1.5: Unlocking multimodal
  understanding across millions of tokens of context,'' \emph{arXiv preprint
  arXiv:2403.05530}, 2024.

\bibitem{park2024utilizingttssynthesizeddata}
H.~J. Park, D.~Agarwal, N.~Chen, R.~Sun, K.~Partridge, J.~Chen, H.~Zhang,
  P.~Zhu, J.~Bartel, K.~Kastner \emph{et~al.}, ``Utilizing {TTS} synthesized
  data for efficient development of keyword spotting model,'' \emph{arXiv
  preprint arXiv:2407.18879}, 2024.

\bibitem{mazumder2021multilingual}
M.~Mazumder, S.~Chitlangia, C.~Banbury, Y.~Kang, J.~M. Ciro, K.~Achorn,
  D.~Galvez, M.~Sabini, P.~Mattson, D.~Kanter \emph{et~al.}, ``Multilingual
  spoken words corpus,'' in \emph{Thirty-fifth Conference on Neural Information
  Processing Systems Datasets and Benchmarks Track (Round 2)}, 2021.

\bibitem{Shen2019LingvoAM}
J.~Shen, P.~Nguyen, Y.~Wu, Z.~Chen, M.~X. Chen, Y.~Jia, A.~Kannan, T.~Sainath,
  Y.~Cao, C.-C. Chiu \emph{et~al.}, ``Lingvo: a modular and scalable framework
  for sequence-to-sequence modeling,'' \emph{arXiv preprint arXiv:1902.08295},
  2019.

\bibitem{warden2018speech}
P.~Warden, ``Speech commands: A dataset for limited-vocabulary speech
  recognition,'' \emph{arXiv preprint arXiv:1804.03209}, 2018.

\bibitem{shin2022learningaudiotextagreementopenvocabulary}
H.-K. Shin, H.~Han, D.~Kim, S.-W. Chung, and H.-G. Kang, ``Learning audio-text
  agreement for open-vocabulary keyword spotting,'' \emph{arXiv preprint
  arXiv:2206.15400}, 2022.

\bibitem{park2024adversarialtrainingkeywordspotting}
H.~J. Park, D.~Agarwal, N.~Chen, R.~Sun, K.~Partridge, J.~Chen, H.~Zhang,
  P.~Zhu, J.~Bartel, K.~Kastner \emph{et~al.}, ``Adversarial training of
  keyword spotting to minimize {TTS} data overfitting,'' \emph{arXiv preprint
  arXiv:2408.10463}, 2024.

\end{thebibliography}

\end{document}